\documentstyle[preprint,epsfig,aps]{revtex}
\tightenlines
\begin{document}
\preprint{\parbox{6cm}{\flushright CLNS 99/1605\\[1cm]}}
\title{Bounding the penguin effects in determinations of $\alpha$ from
$B^0(t)\to \pi^+\pi^-$}
\author{Dan Pirjol}
\address{Floyd R. Newman Laboratory of Nuclear Studies,\\
Cornell University, Ithaca, New York 14853 }
\date{\today}
\maketitle
\begin{abstract}
In the absence of the QCD penguin contributions a measurement of the time-dependent
asymmetry in the decay $B^0(t)\to \pi^+\pi^-$ gives directly the weak angle $\alpha$.
Several bounds have been proposed in the literature on the magnitude of the penguin
effects on this determination, the prototype of which is the isospin bound of
Grossman and Quinn. We point out that large strong final state interactions could 
cause these bounds to overestimate the real penguin effect.
A new flavor SU(3) bound is proposed, requiring  only the 
charge-averaged branching ratios for $B^0\to \pi^+\pi^-$ and $B_s\to K^+ K^-$, which 
exactly takes into account all relevant amplitudes and electroweak penguin effects. 
This bound on the penguin-induced error on the determination of the weak phase $\alpha$ 
holds even without a knowledge of the direct CP asymmetry in the $\pi^+\pi^-$ channel.
\end{abstract}
\pacs{pacs1,pacs2,pacs3}

\narrowtext

{\bf 1.} The determination of the weak angle $\alpha$ from time-dependent measurements 
of the
decay $B^0(t)\to\pi^+\pi^-$ is afflicted with theoretical uncertainties due to the
penguin contamination (see e.g. \cite{Fleisch,BaBar}). In principle this uncertainty can
be resolved through the isospin construction of Gronau and London \cite{GroLo}, which
involves measuring the separate rates for $B^0\to\pi^0\pi^0$ and its charge-conjugate.
Unfortunately, experimental difficulties with tagging decay modes of neutral B mesons
into $\pi^0\pi^0$ will limit the precision of this method \cite{BaBar}.

Several other approaches have been proposed to circumvent this problem \cite{PT}.
Recently Grossman and Quinn \cite{GroQu} presented a simple method for bounding the effect
of the penguins on the determination of $\alpha$, using only the
charge-averaged rate for $B^0\to\pi^0\pi^0$ (and $B^\pm\to\pi^\pm\pi^0$), which would be 
easier to measure than the separate rates for $B^0$ and $\bar B^0$. 
This bound has been improved in \cite{Charles}, where yet another isospin bound has been
proposed. 
In addition, the paper \cite{Charles} presents several bounds based on flavor SU(3) 
symmetry,  together with dynamical assumptions about the size of certain OZI-suppressed
penguin contributions. While less rigorous, these bounds turn out to be more
informative than the isospin bounds when combined with recent CLEO data on 
$B\to K\pi$ decays.

In this note we point out that the isospin bounds of Grossman-Quinn \cite{GroQu}
and Charles \cite{Charles} are only likely to offer useful information on the penguin 
contamination if the tree-level contributions to the $B^0\to \pi^0\pi^0$ amplitude
are small. Unfortunately, they are not necessarily so and can conceivably be enhanced 
by final state interactions \cite{rescatt}. This raises the concern that the
resulting bounds might have little bearing on the size of the penguin contamination
and measure instead the size of the rescattering effects. We present an improved
bound which takes into account all relevant amplitudes and is based only on flavor SU(3)
symmetry. This bound requires only the time-independent charge-averaged rates for $B^0\to
\pi^+\pi^-$ and $B_s\to K^+K^-$, which are easier to measure than the rate into two
neutral pions. At the same time, the theoretical status of this bound is cleaner
than that of the other SU(3) bounds proposed in \cite{Charles}.

\vspace{5mm}

{\bf 2.} The angle $\alpha$ can be measured from the time-dependence of the $CP$-asymmetry
\begin{eqnarray}\nonumber
a_{+-}(t) &\equiv& 
\frac{|\langle\pi^+\pi^-|B^0(t)\rangle|^2 - |\langle\pi^+\pi^-|\bar B^0(t)\rangle|^2}
{|\langle\pi^+\pi^-|B^0(t)\rangle|^2 + |\langle\pi^+\pi^-|\bar B^0(t)\rangle|^2}\\
\label{1}
&=& a_{\rm dir} \cos(\Delta mt) + \sqrt{1-a^2_{\rm dir}}
\sin(2\alpha+2\theta)\sin(\Delta mt)\,,
\end{eqnarray}
where $\Delta m$ is the mass difference of the eigenstates of the $B^0-\bar B^0$ system.
(See \cite{BuFl} for an alternative determination.) The strength of the direct CP violation 
in $B^0\to \pi^+\pi^-$ is measured by the parameter
\begin{eqnarray}
a_{\rm dir} \equiv \frac{|A^{+-}|^2-|\bar A^{+-}|^2}{|A^{+-}|^2+|\bar A^{+-}|^2}\,.
\end{eqnarray}
Using the notations of \cite{GPY} the amplitudes $A^{+-}$ and $\bar A^{+-}$ are written as
follows in terms of graphical SU(3) amplitudes
\begin{eqnarray}
A^{+-} &\equiv& A(B^0\to\pi^+\pi^-)\nonumber\\\label{2}
 & & = -|\lambda_u^{(d)}|e^{i\gamma}(T+E+P_{uc}+PA_{uc}+P^{EW}_{uc})
- |\lambda_t^{(d)}|e^{-i\beta}(P_{ct}+PA_{ct}+P^{EW}_{ct})\\
\bar A^{+-} &\equiv& A(\bar B^0\to\pi^+\pi^-)\nonumber\\\label{3}
 & & = -|\lambda_u^{(d)}|e^{-i\gamma}(T+E+P_{uc}+PA_{uc}+P^{EW}_{uc})
- |\lambda_t^{(d)}|e^{i\beta}(P_{ct}+PA_{ct}+P^{EW}_{ct})\,.
\end{eqnarray}
The angle $\theta$ appearing in (\ref{1}) is given by $2\theta=$ Arg$(\tilde A^{+-}/A^{+-})$, 
with $\tilde A^{+-}\equiv e^{2i\gamma}\bar A^{+-}$. One can see that, in the absence of the 
second terms
in (\ref{2}), (\ref{3}) (no penguin contamination), this angle vanishes and $\alpha$ can
be simply extracted from the $\sin(\Delta mt)$ part of the $CP$-asymmetry (\ref{1}).
Our problem in the following will be to set an upper bound on the angle $\theta$.

Starting from the graphical construction of Gronau and London \cite{GroLo}, 
Grossman and Quinn derived the following upper bound on $\theta$
\begin{eqnarray}\label{4}
\sin^2 \theta \leq R_{\pi^0\pi^0}\equiv \frac{<\mbox{B}(B^0\to\pi^0\pi^0)>}
{<\mbox{B}(B^\pm\to\pi^\pm\pi^0)>}
\end{eqnarray}
where the angular brackets stand for flavor-averaged decay rates. We will briefly review
in the following their argument.

The angle $\theta$ can be determined from the isospin triangles (the decay amplitudes are
identified by the charge of the two pions in the final state) \cite{GroLo}
\begin{eqnarray}\label{5}
A^{+-} + \sqrt2 A^{00} = \sqrt2 A^{+0}\,,\qquad
\tilde A^{+-} + \sqrt2 \tilde A^{00} = \sqrt2 \tilde A^{-0}\,.
\end{eqnarray}
Furthermore, neglecting small (and calculable)
electroweak penguin contributions \cite{BuFle,GPY}, one has the equality $A^{+0} =
\tilde A^{-0}$. This fixes the relative orientation of the two isospin triangles,
such that the angle $2\theta$ between $A^{+-}$ and $\tilde A^{+-}$ can be directly extracted
from a knowledge of the rates for the 5 processes in (\ref{5}). There are discrete 
ambiguities in this determination, corresponding to the freedom of drawing the two
triangles (\ref{5}) on the same or on opposite sides of the common side $A^{+0}$.
The maximal value of $\theta$ corresponds, obviously, to the latter possibility,
which will be implicitly understood in the following.

Let us keep first $|A^{00}|, |\tilde A^{00}|$ and $|A^{+0}|$ fixed. Simple geometrical
considerations show that the angle $2\theta=$Arg$(\tilde A^{+-}/A^{+-})$ is maximized
when each of $\theta_1=$Arg$(\tilde A^{+-}/A^{+0})$ and $\theta_2=$Arg$(A^{+0}/\tilde A^{+-})$
are separately maximized. Their respective maximum values are
given by $\sin(\theta_1)_{\rm max}=|A^{00}|/|A^{+0}|$ and 
$\sin(\theta_2)_{\rm max}=|\tilde A^{00}|/|A^{+0}|$. Next, one must find the maximum value of
$2\theta=\theta_1+\theta_2$ for all values of $|A^{00}|, |\tilde A^{00}|$ such that
$|A^{00}|^2+ |\tilde A^{00}|^2=\langle |A^{00}|^2\rangle$ is fixed. 
The extremum value of $\theta$ is easily found as
\begin{eqnarray}\nonumber
\mbox{min}(\cos(2\theta)) &=& \mbox{min}\left\{\sqrt{\left(1-\frac{|A^{00}|^2}{|A^{+0}|^2}\right)
\left(1-\frac{|\tilde A^{00}|^2}{|A^{+0}|^2}\right)} - \frac{|A^{00}|}{|A^{+0}|}
\frac{|\tilde A^{00}|}{|A^{+0}|}\right\}\\
\label{6}
&=& \mbox{min}\left\{\sqrt{1 - 2R_{\pi^0\pi^0} + x^2} - x\right\}\,,\qquad
x\equiv \frac{|A^{00}|}{|A^{+0}|}\frac{|\tilde A^{00}|}{|A^{+0}|}
\end{eqnarray}
with the constraint
\begin{eqnarray}\label{7}
\frac{|A^{00}|^2}{|A^{+0}|^2}+
\frac{|\tilde A^{00}|^2}{|A^{+0}|^2}=2R_{\pi^0\pi^0}\,.
\end{eqnarray}
The function $f(x)\equiv \sqrt{1-a^2+x^2}-x$ is uniformly decreasing such that the minimum
of the expression (\ref{6}) is attained for the maximum allowed value of $x$ compatible with
the constraint (\ref{7}). This is reached at $x=R_{\pi^0\pi^0}$ and corresponds to the case 
$|A^{00}|=
|\tilde A^{00}|$, or $\theta_1=\theta_2=\theta$, which justifies the inequality (\ref{4}).

This bound has been improved in \cite{Charles} under the assumption that the direct
CP asymmetry $a_{\rm dir}$ in $B^0\to\pi^+\pi^-$ is known:
\begin{eqnarray}
\cos(2\theta_{\rm max}) = \frac{1}{\sqrt{1-a^2_{\rm dir}}}
\left(1 -
2\frac{<\mbox{B}(B^0\to\pi^0\pi^0)>}
{<\mbox{B}(B^\pm\to\pi^\pm\pi^0)>}\right)\,.
\end{eqnarray}
Another isospin bound has been given in \cite{Charles}, in terms of the ratio
$\mbox{B}(B^0\to\pi^0\pi^0)/\mbox{B}(B^0\to\pi^+\pi^-)$.

How effective can one expect these results to be in controlling the penguin effect in
(\ref{2}), (\ref{3})? The isospin bounds are formulated in terms of the amplitude of the
decay $B^0\to\pi^0\pi^0$ relative to $A(B^\pm\to\pi^\pm\pi^0)$ (or $A(B^0\to\pi^+\pi^-)$). 
However, an examination of the former amplitude shows that it contains several components 
as well, in addition to the penguin
\begin{eqnarray}\label{8}
\sqrt2 A^{00} &\equiv& \sqrt2 A(B^0\to\pi^0\pi^0) = 
-|\lambda_u^{(d)}|e^{i\gamma}(C-E-P_{uc}-PA_{uc})
+ |\lambda_t^{(d)}|e^{-i\beta}(P_{ct}+PA_{ct})\\
\sqrt2 A^{+0} &\equiv& \sqrt2 A(B^0\to\pi^+\pi^0) = 
-|\lambda_u^{(d)}|e^{i\gamma}(T+C)\,.
\end{eqnarray}
The small contributions of electroweak penguins was neglected here.
The amplitude (\ref{8}) satisfies the triangle inequality $|(-C+E) + (P+PA)|\geq
|P+PA| - |C-E|$, where we assumed that the penguin amplitude dominates over the tree
contribution $C-E$. This inequality is easily translated into an upper bound on the
magnitude of the penguin amplitude $P+PA$, which is the essence of the Grossman-Quinn
bound.
However, this bound will only be useful if the amplitude (\ref{8}) is indeed dominated
by the penguin amplitude. If this is not the case, the bound will measure instead the
size of the amplitude $C-E$ and will have little bearing on the penguin contribution.

From (\ref{8}) one can see that, even if the penguin contribution vanished,
the ratio $R_{\pi^0\pi^0}$ would not vanish together with it, but would take a finite value
of the order $|(C-E)/(T+C)|^2$. Naive arguments based on the factorization approximation
suggest that this ratio is small, of the order $R_{\pi^0\pi^0}\simeq (0.2)^2$ \cite{GLHR}, 
which implies values for $\theta$ around $11^\circ$. However, rescattering
effects can shift the precise numerical value through the exchange amplitude $E$ \cite{rescatt} 
in an 
unknown direction. The potentially large value of the ratio $|(C-E)/(T+C)|^2$ could
therefore diminish the usefulness of the bound (\ref{4}). We stress that even in this
case the bound  will continue to hold; however, it will most likely
overestimate  the real  penguin effect\footnote{
A different source of corrections to the Grossman-Quinn bound is due to isospin breaking 
effects. They were studied in \cite{Gardner} and shown to the rather small.}.\vspace{5mm}

{\bf 3.} 
The magnitude of the angle $\theta$ is closely related to the ratio of the components
of different weak phases in the amplitude $A^{+-}(\tilde A^{+-})$:
\begin{eqnarray}\label{10prev}
A^{+-} \propto e^{i\gamma} + r e^{i\phi}\,,\qquad
\tilde A^{+-} \propto e^{i\gamma} + r e^{i(\phi+2\gamma)}\,.
\end{eqnarray}
We made use of the unitarity of the CKM matrix to write $\lambda_t^{(s)}=-\lambda_u^{(s)}
-\lambda_c^{(s)}$. Explicitly, the parameters $r, \phi$ are given by
\begin{eqnarray}
re^{i\phi}\equiv \frac{1}{R_b(1-\lambda^2/2)} 
\frac{P_{ct}+PA_{ct}+P_{ct}^{EW}}
{(T+E+P_{uc}+PA_{uc}+P^{EW}_{uc})-(P_{ct}+PA_{ct}+P^{EW}_{ct})}\,.
\end{eqnarray}
We denoted here as usual $R_b=1/\lambda |V_{ub}/V_{cb}|$ with $\lambda=0.22$ one of the
Wolfenstein parameters.

We will show now that very useful constraints on $r$ and ultimately
on $\theta$ can be obtained from a consideration of the $\Delta S=1$ decay 
$B_s\to K^+ K^-$. This is the main $B_s$ decay mode with an expected branching ratio at the
${\cal O}(10^{-5})$ level which will be easily measured at the hadronic machines. 
Its decay amplitude is given by
\begin{eqnarray}\label{10}
A(B_s\to K^+ K^-) = -|\lambda_u^{(s)}|e^{i\gamma}(T+E+P_{uc}+PA_{uc}+P^{EW}_{uc})
- \lambda_t^{(s)}(P_{ct}+PA_{ct}+P^{EW}_{ct})\,.
\end{eqnarray}
In the limit of SU(3) symmetry, exactly the same amplitudes enter both (\ref{2}) 
and (\ref{10}). This fact can be used to constrain the ratio of the two amplitudes
with different weak phase, in analogy to a similar bound on the annihilation
amplitude in $B^\pm\to K^0\pi^\pm$ \cite{Falk,Limits,Neubert}.
In the normalization of (\ref{10prev}), the amplitudes for $B_s\to K^+ K^-$ and its
charge conjugate are given by
\begin{eqnarray}\label{Bsampl}
A_{K^+ K^-}\propto \lambda'
e^{i\gamma} - \frac{1}{\lambda'} r e^{i\phi}\,,\quad
\tilde A_{K^+ K^-}\propto \lambda'
e^{i\gamma} - \frac{1}{\lambda'} r e^{i(\phi+2\gamma)}\,.
\end{eqnarray}
We denote here and in the following $\lambda'=\lambda/(1-\lambda^2/2)$.

\begin{figure}[hhh]
 \begin{center}
 \mbox{\epsfig{file=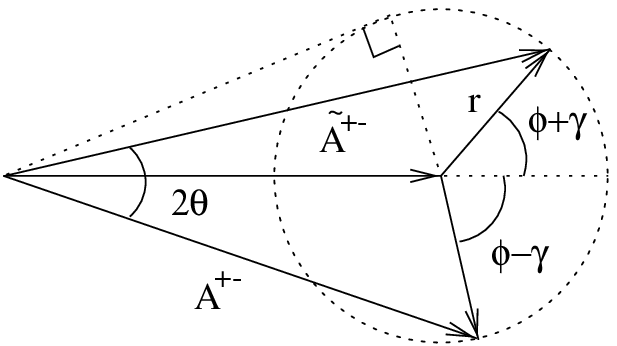,width=10cm}}
 \end{center}
 \caption{
Graphical representation of the amplitudes $A^{+-}\,,\tilde A^{+-}$ (\ref{2}), 
(\ref{3}). For given radius of the circle $r$ (``penguin/tree'' ratio)
the maximum value of $\theta$
corresponds to the dotted line and is given by $\sin\theta_{\rm max}=r$.
}
\label{fig1}
\end{figure}

Let us introduce the ratio of charge-averaged decay rates
\begin{eqnarray}\label{R+-}
R_{K^+ K^-}\equiv \frac{\langle \mbox{B}(B_s\to K^\pm K^\mp)\rangle}
{\langle \mbox{B}(B^0\to\pi^\pm\pi^\mp)\rangle} =
\frac{\displaystyle \lambda'^2 +
\frac{1}{\lambda'^2} r^2 - 2r\cos\gamma\cos\phi}
{\displaystyle 1 + r^2 + 2r\cos\gamma\cos\phi}\,,
\end{eqnarray}
Varying $\gamma$
and $\phi$ through all their possible values, one finds that the $r$ parameter
(the ``penguin/tree'' ratio) is constrained by the inequalities
\begin{eqnarray}
\frac{\displaystyle\sqrt{R_{K^+ K^-}}-\lambda'}
{\displaystyle\frac{1}{\lambda'} + \sqrt{R_{K^+ K^-}}} \equiv r_{\rm min}
\leq r \leq 
r_{\rm max}\equiv \frac{\displaystyle\sqrt{R_{K^+ K^-}}+\lambda'}
{\displaystyle\frac{1}{\lambda'} - \sqrt{R_{K^+ K^-}}}\,,
\end{eqnarray}
At first sight, the upper bound on $r$ together with simple geometrical considerations 
(see Fig.~\ref{fig1}) would lead one to assume a similar bound on $\theta$ given by 
$\sin\theta\leq  r_{\rm max}$.
In fact, due to a correlation in $\phi$ between $\theta$ and $R_{K^+ K^-}$, the correct
bound is more restrictive than this. To find it, let us fix $\gamma$ 
and vary $\phi$ (at fixed $R_{K^+ K^-}$) such as to maximize $\theta$. 
The maximum value of $\theta$ is given by
\begin{eqnarray}\label{Rreal}
\sin\theta_{\rm max}(R_{K^+ K^-},\gamma) = \frac{r(R_{K^+ K^-},\gamma)\sin\gamma}
{\sqrt{1+r^2(R_{K^+ K^-},\gamma)\mp 2r(R_{K^+ K^-},\gamma)\cos\gamma}}
\end{eqnarray}
where the upper (lower) sign corresponds to $0\leq \gamma\leq \pi/2$ 
($\pi/2\leq \gamma\leq \pi$), and  is reached at  $\phi=0\, (\pi)$.
In this formula $r=r(R_{K^+ K^-},\gamma)$ is to be obtained from (\ref{R+-}) 
after substituting the  correct value for $\phi$ for each given value of $\gamma$,
as explained above.

\begin{figure}[hhh]
 \begin{center}
 \mbox{\epsfig{file=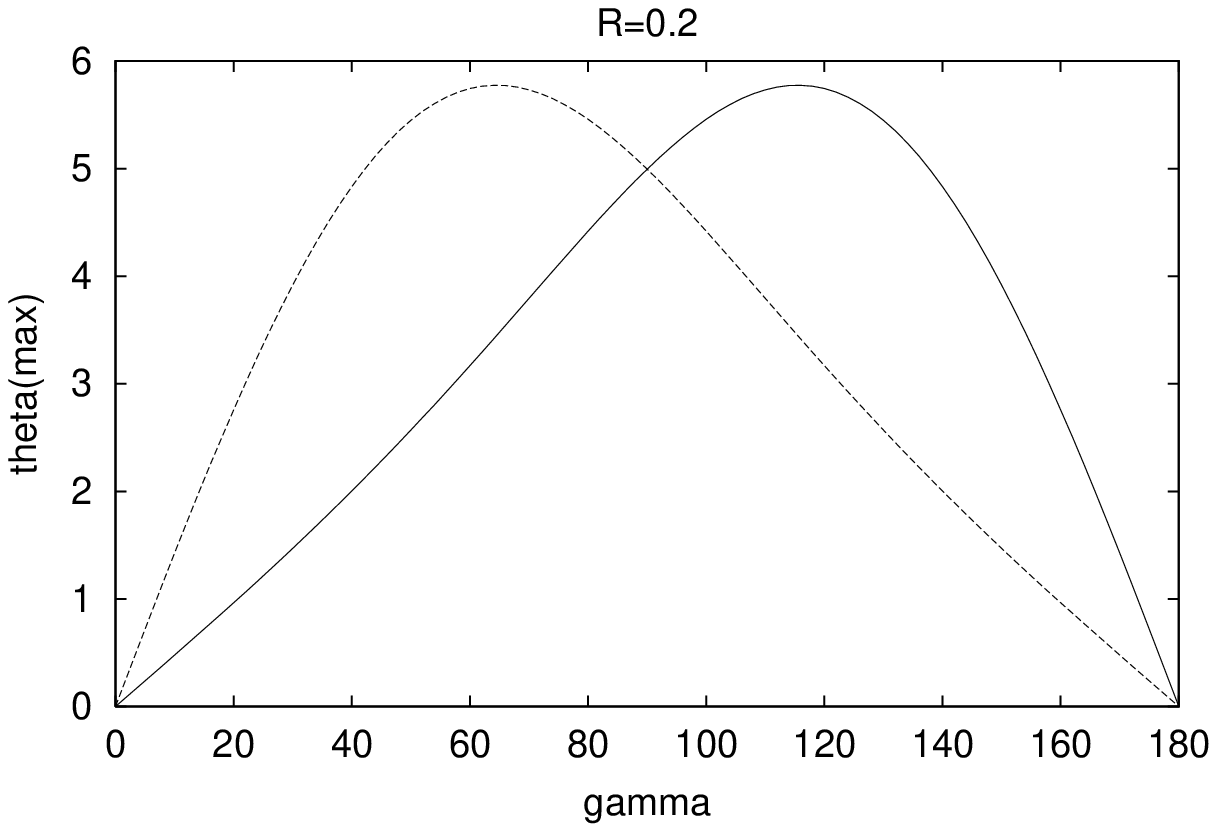,width=8cm}}
 \mbox{\epsfig{file=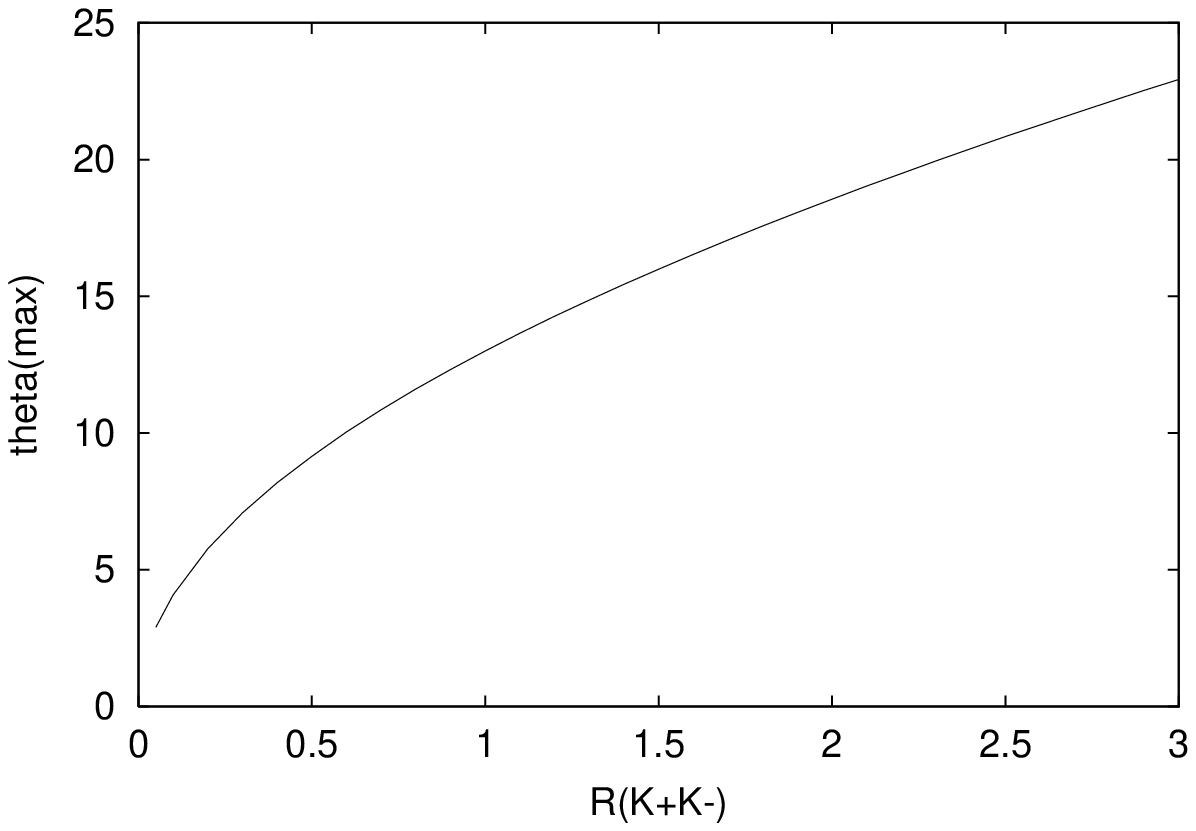,width=8cm}}
 \end{center}
 \caption{
Left: Plot of $\theta_{\rm max}(\gamma)$ $[^\circ ]$ for fixed $R_{K^+ K^-}=0.2$;
the solid (dotted) line corresponds to $\phi=\pi$ $(0)$. Right: 
$\theta_{\rm max}(R_{K^+ K^-})$ maximized over all values of $\gamma$, as a function
of the ratio $R_{K^+ K^-}$.
}
\label{fig2}
\end{figure}

Typical numerical results for $\theta_{\rm max}(\gamma)$ are shown in Fig.~\ref{fig2} 
for one value of $R_{K^+ K^-}$. This angle reaches a maximum at two intermediate
values of $\gamma$, symmetrical with respect to 90$^\circ$, approaching this
value in the limit of a large $R_{K^+ K^-}$. The resulting bound,
maximized over $\gamma$, is shown in Fig.~\ref{fig2} as a function of the ratio
$R_{K^+ K^-}$; this is about a factor of two more restrictive than
the naive bound $\sin\theta_{\rm max}=r_{\rm max}$. 

The bound on $\theta_{\rm max}(R_{K+ K-})$ can be slightly improved if some information 
on $\gamma$ is available. This would be possible provided that $\gamma$ can be restricted to
a sufficiently narrow region around 90$^\circ$ (see Fig.~\ref{fig2}). The present constraint
$0.54\leq \sin^2\gamma\leq 1$ from a global analysis of the unitarity triangle \cite{JRos} 
is not restrictive enough to allow an improvement for any value of $R_{K+ K-}$.

The results obtained so far assume that the direct CP violation parameter $a_{\rm dir}$
in (\ref{1}) is known. We will show next that in fact they do not require the knowledge of
this parameter, just as the bounds in \cite{GroQu,Charles}. This is useful in 
practice since
the error in the determination of $a_{\rm dir}$ will not propagate into the final error on 
$\alpha$. We will prove in the following the stronger constraint
\begin{eqnarray}\label{Delta}
\sin(2\alpha - 2\bar\theta_{\rm max}) &\leq&
\sin(2\alpha + \psi - |2\theta_{\rm max}-\psi|) \leq
\sqrt{1-a^2_{\rm dir}}\sin(2\alpha + 2\theta) \nonumber\\
& &\leq \sin(2\alpha + \psi + |2\theta_{\rm max}-\psi|) \leq
\sin(2\alpha + 2\bar\theta_{\rm max})\,,
\end{eqnarray}
for {\em any} value of $\alpha$ (not necessarily identical to the weak phase). 
The angle $\psi(\gamma)$ is defined below in (\ref{21}) and we denoted with
$\bar\theta_{\rm max}$ the value of $|\theta_{\rm max}|$ maximized over all
values of $\gamma$ (plotted in Fig.~2 (right)).
In this expression both $a_{\rm dir}$ and $\theta$ are functions of the unknown
strong phase $\phi$, which is taken completely arbitrary in the interval $(0,2\pi)$.
In the notations of (\ref{10prev}),  the $a_{\rm dir}$ parameter is given by
\begin{eqnarray}
a_{\rm dir} = \frac{2r\sin\gamma\sin\phi}{1+r^2+2r\cos\gamma\cos\phi}\,,
\end{eqnarray}
with $r(R_{K^+ K^-},\gamma,\phi)$ determined from (\ref{R+-}).
The correlation between $\theta$ and $a_{\rm dir}$ at fixed $R_{K^+ K^-}$ and $\gamma$
is made explicit  by the following relation
\begin{eqnarray}\label{20}
R_{K^+ K^-} = \frac{\displaystyle \lambda'+\frac{1}{\lambda'}}{2\sin^2\gamma}\left\{
\lambda'+\frac{1}{\lambda'} - \sqrt{1-a^2_{\rm dir}}\left[
\lambda'\cos(2\theta-2\gamma) + \frac{1}{\lambda'}\cos(2\theta)\right]\right\} -1 \,.
\end{eqnarray}
This can be proven by expressing $r^{i\phi}$ in (\ref{10prev}) in terms of
$A^{+-}, \tilde A^{+-}$ and substituting the result into (\ref{Bsampl}). 
From (\ref{20}) one finds the following very useful relation at fixed $R_{K^+ K^-}$ and
$\gamma$
\begin{eqnarray}\label{21}
& &\sqrt{1-a^2_{\rm dir}} \cos(2\theta - \psi(\gamma)) = \mbox{Ct.} = 
\cos(2\theta_{\rm max} - \psi(\gamma))\quad (0<\mbox{Ct.}\leq 1)\,,\nonumber\\
& &\qquad\qquad\tan\psi(\gamma) \equiv \frac{\lambda'\sin(2\gamma)}{\displaystyle 
\lambda'\cos(2\gamma) + \frac{1}{\lambda'}}\,. 
\end{eqnarray}
From this relation it is easy to read off the range of variation of the parameters 
$\theta$ and $a_{\rm dir}$. The maximum value of $\theta$ is reached for $a_{\rm dir}=0$,
and is given by $2\theta_{\rm max}=\pm|\arccos(\mbox{Ct.})|+\psi(\gamma)$. The sign of the
$\arccos$ must be chosen opposite to that of $\psi(\gamma)$ (such as to maximize $|\theta|$)
which is in turn given by $\psi(\gamma) > 0 (<0)$, for $0\leq \gamma\leq \pi/2$ 
($\pi/2\leq \gamma\leq \pi$). One obtains $\theta_{\rm max} >0 (<0)$ for these two
cases respectively, in agreement with (\ref{Rreal}). From this relation it is also clear that 
$|2\theta-\psi(\gamma))| < \pi/2$.

We are now ready to prove the inequality stated in (\ref{Delta}). This is
easily done by writing
\begin{eqnarray}
\sqrt{1-a^2_{\rm dir}}\sin(2\alpha+2\theta) = 
\sin(2\alpha+\psi)\cos(2\theta_{\rm max}-\psi) +
\sqrt{1-a^2_{\rm dir}}\cos(2\alpha+\psi)\sin(2\theta-\psi)
\end{eqnarray}
and using $-\sin|2\theta_{\rm max}-\psi| \leq\sqrt{1-a^2_{\rm dir}}\sin(2\theta-\psi)\leq
\sin|2\theta_{\rm max}-\psi|$ in the second term.

One notes from (\ref{20}) that the bound on $\theta$ can be improved if the direct CP 
asymmetry parameter $a_{\rm dir}$ is known. One obtains in this case
\begin{eqnarray}
\cos(2\theta_{\rm max}-\psi(\gamma)) = \frac{1}{\sqrt{1-a^2_{\rm dir}}}
\left(
\frac{\displaystyle \left(\lambda'+\frac{1}{\lambda'}\right)^2-2\sin^2\gamma (1+R_{K^+ K^-})}
{\displaystyle\left(\lambda'+\frac{1}{\lambda'}\right)\sqrt{\left(\lambda'+
\frac{1}{\lambda'}\right)^2-4\sin^2\gamma}} \right)\,.
\end{eqnarray}
Furthermore, if in addition also the weak angle $\gamma$ is known, then the 
 constraint on $\theta$  turns  into a determination of this angle.
In practice however it is more likely
that only a constraint will be available on the latter. We will not pursue this further
and ask instead what constraints can be set on $a_{\rm dir}$ once the value of $R_{K^+ K^-}$
is measured.
As already mentioned, the minimum value which can be taken by $a_{\rm dir}$ is 0
(reached at $\theta=\theta_{\rm max}, -\theta_{\rm max}+\psi$). The maximum value is
reached at $\theta=\psi(\gamma)/2$ and is given by
\begin{eqnarray}
& &|a_{\rm dir}|_{\rm max} = \frac{2|\sin\gamma|}{\displaystyle\lambda'+\frac{1}{\lambda'}}
\sqrt{
\frac
{\displaystyle R_{K^+ K^-}\left(\lambda'+\frac{1}{\lambda'}\right)^2-(1+R_{K^+ K^-})^2\sin^2\gamma}
{\displaystyle\left(\lambda'+ \frac{1}{\lambda'}\right)^2-4\sin^2\gamma}}\nonumber\\
& & \qquad\qquad \leq
\frac{2}{\displaystyle\frac{1}{\lambda'^2}-\lambda'^2}
\sqrt{R_{K^+ K^-}\left(\lambda'+\frac{1}{\lambda'}\right)^2-(1+R_{K^+ K^-})^2}\,.
\end{eqnarray}

So far we neglected SU(3) breaking effects, which can be expected to dominate the
theoretical error of this method. Some insight into the size of these effects can be 
obtained as follows. In the SU(3) limit the CP asymmetries in $B^0\to \pi^+\pi^-$ and 
$B_s\to K^+ K^-$ are related in a simple way to the ratio $R_{K^+ K^-}$
\begin{equation}\label{SU3breaking}
\frac{a_{\rm dir}^{\pi^+\pi^-}}{a_{\rm dir}^{K^+K^-}} = -R_{K^+ K^-}\,.
\end{equation}
The amount of violation of this relation could be used as a measure of the SU(3) breaking
effects in the bound on $\theta$.

\vspace{5mm}
{\bf 4.} 
We turn now to a numerical discussion of the bound on $\theta$.
In the flavor SU(3) and zero penguin limit, the ratio
(\ref{R+-}) takes the value $[R_{K^+ K^-}]_0=\lambda'^2$, which implies
 $\theta_{\rm max} \simeq 2.9^\circ$. This is much more constraining,
by almost a factor of four, than the naive estimate of the Grossman-Quinn bound (\ref{4})
taken in the same limit.

The real question is, of course, how large the ratio $R_{K^+ K^-}$ can be in the
presence of the penguin amplitude? A simple estimate may be given by assuming that the
annihilation-topology penguin amplitude $PA$ is negligible. Such an approximation can well
turn out to be very wrong, as this amplitude can receive large strong interaction final 
state effects \cite{rescatt}. Still, it is useful as a rough order of magnitude estimate.

Neglecting furthermore the tree contribution in (\ref{10}) and assuming flavor SU(3) 
symmetry, this amplitude can be related to the amplitude for $B^\pm\to K^0\pi^\pm$
for which preliminary CLEO data is available.
This gives $|A(B_s\to K^+ K^-)|\simeq |A(B^+\to K^0\pi^+)|$, which translates into
\begin{eqnarray}\label{number}
R_{K^+ K^-} \simeq \frac{\langle \mbox{B}(B^\pm\to K^0 \pi^\pm)\rangle}
{\langle \mbox{B}(B^0\to\pi^\pm\pi^\mp)\rangle}  \simeq 
1.67 - 3.5\,.
\end{eqnarray}
We used here the CLEO data $B(B^\pm\to K^0\pi^\pm)=(1.4\pm 0.5\pm 0.2)\times 10^{-5}$
and $0.4\times 10^{-5} < B(B^0\to \pi^\pm\pi^\mp) < 0.84\times 10^{-5}$ \cite{CLEO} 
(the lower bound on the latter is marginally compatible with the CLEO preliminary
result $B(B^0\to \pi^\pm\pi^\mp)_{2.9\sigma} = (0.37^{+0.20}_{-0.17})\times 10^{-5}$).
From Fig.~\ref{fig2} the result (\ref{number}) implies an upper limit on $\theta$ 
of 15 -- 25$^\circ$, which agrees with other estimates with a similar theoretical
input \cite{Charles} (SU(3) symmetry and neglect of annihilation topologies).
The maximum value of $a_{\rm dir}$ corresponding to this range is $a_{\rm dir}=0.55-0.76$.

 Of course, it remains to be seen how the annihilation-topology penguin amplitude
$PA$ will modify this prediction. If its effect is a suppression of the ratio $R_{K^+ K^-}$
then the bound on $\theta$ will be correspondingly tightened.

To summarize the points of our discussion:

(1) The bound (\ref{Rreal}) on $\theta=\alpha_{\rm eff}-\alpha$ requires only the
easily accessible $B$ decay mode $B_s\to K^+ K^-$. Furthermore, no tagging is required,
as only charge-averaged rates enter this relation. The bound does not require the
knowledge of the CP asymmetry parameter $a_{\rm dir}$.

(2) No assumption beyond flavor SU(3) symmetry is required. In particular,
electroweak penguin  contributions and the annihilation-type
penguin amplitude $PA$ are taken into account exactly. To our knowledge, aside from the
isospin bounds (\cite{GroQu}, \cite{Charles}), this is the only existing bound holding 
with these minimal assumptions. (A bound proposed in \cite{Charles} making use
of the rate $B(B^0\to K^0\bar K^0)$ assumes the vanishing of the penguins
with internal charm quarks and neglects electroweak penguin effects.) 

(3) On the down side, the use of SU(3) symmetry introduces additional uncertainties
in the bound on $\theta$. Some control on the size of these effects could be obtained 
phenomenologically from the violation of the SU(3) relation (\ref{SU3breaking}).

\vspace{5mm}
{\em Acknowledgements.} This work has been supported by the National Science
Foundation. I am grateful to Michael Gronau and Yuval Grossman for discussions and 
comments on the manuscript.

\newpage
\thispagestyle{plain}

\end{document}